\documentclass[twocolumn,aps,prd,groupedaddress,showpacs]
              {revtex4}%
\usepackage{graphicx}
\usepackage{amsmath}
\usepackage{bm}
\usepackage{relsize}
\RequirePackage{xspace}
\newcommand{\bqa}{\begin{eqnarray}}
\newcommand{\eqa}{\end{eqnarray}}
\newcommand{\pslash}{\slash\hspace{-0.55em}}
\newcommand{\as}{\alpha_{\mathrm{s}}}

\begin{document}


\title{\mbox{}\\[10pt]
$B\to\chi_{c1}(1P,2P)K$ decays in QCD factorization and X(3872)}

\author{Ce Meng$~^{(a)}$, Ying-Jia Gao$~^{(a)}$, and Kuang-Ta Chao$~^{(b,a)}$}
\affiliation{ {\footnotesize (a)~Department of Physics, Peking
University,
 Beijing 100871, People's Republic of China}\\
{\footnotesize (b)~China Center of Advanced Science and Technology
(World Laboratory), Beijing 100080, People's Republic of China}}




\begin{abstract}
$B\to\chi_{c1}(1P,2P)K$ decays are studied in QCD factorization by
treating charmonia as nonrelativistic bound states. No infrared
divergences exist in the vertex corrections, while the logarithmic
end-point singularities in the hard spectator corrections can be
regularized by a momentum cutoff. Within certain uncertainties we
find that the $B\to\chi_{c1}(2P)K$ decay rate can be comparable to
$B\to\chi_{c1}(1P)K$, and get $Br(B^0 \!\rightarrow \!\chi_{c1}'
K^0)\! =\!Br(B^+ \!\rightarrow \!\chi_{c1}' K^+)\!\approx\!
2\times 10^{-4}$. This might imply a possible interpretation for
the newly discovered X(3872) that this state has a dominant
$J^{PC} = 1^{++}(2P)$ $c\bar c$ component but mixed with a
substantial $D^0\bar{D}^{*0}+D^{*0}\bar{D}^0$ continuum component.
\end{abstract}


\maketitle



The naively factorizable decay \cite{BSW} $B\rightarrow \chi_{c1}
K$ was studied\cite{Chao03} in the QCD factorization approach
\cite{BBNS} in which the nonfactorizable vertex and spectator
corrections were also estimated, but the numerical results were
four times smaller than experimental data.  Recently, these decays
were also studied in the PQCD approach \cite{Li}.  In both the
above approaches, light-cone distribution amplitudes(LCDAs) were
used to describe $\chi_{c1}$. As argued in Ref.~\cite{Chao04}, a
more appropriate description of charmonium is the nonrelativistic
(NR) wave functions which can be expanded in terms of the relative
momentum $q$ between charm and anticharm quarks.  This argument is
based on the nonrelativistic nature of heavy quarkonium
\cite{BBL1}.  With careful studies, we find that the two
descriptions (i.e.LCDAs and NR) are equivalent for the S-wave
charmonium states (see,e.g. \cite{Chao02}), but in the case of
P-wave states the light-cone descriptions lose some important
contributions in the leading-twist approximation. This is not
surprising since $q$ can be neglected in S-wave states, but cannot
be neglected for P-wave states even in leading order
approximation.

On the phenomenological hand, the study of $B\to\chi_{c1}(2P)K$
may help clarify the nature of the recently discovered resonance
$X(3872)$\cite{Belle03}, since the measurements for $X(3872)$
favor $J^{PC} = 1^{++}$\cite{Belle05} and hence $\chi_{c1}(2P)$
becomes one of the possible assignments for it. On the other hand,
aside from the conventional charmonium\cite{BG,chao1}, a loosely
bound S-wave molecule of $D^0\bar{D}^{*0}+D^{*0}\bar{D}^0$  has
been suggested for X(3872)\cite{Tornqvist,Braaten}.

Motivated by the above considerations, in this paper we study the
decays $B\to\chi_{c1}(1P,2P)K$ within the framework of QCD
factorization by treating the charmonia $\chi_{c1}(1P,2P)$ as
nonrelativistic bound states with $m_c/m_b$ taken to be a fixed
value in the heavy $b$ quark limit. We will estimate the
production rate of $\chi_{c1}(2P)$ and argue that the X(3872) may
be dominated by the $\chi_{c1}(2P)$ charmonium but mixed with some
$D^0\bar{D}^{*0}+D^{*0}\bar{D}^0$ continuum component.

In the non-relativistic bound-state picture, charmonium can be
described by the color-singlet NR wave function.  Let $p$ be the
total momentum of the charmonium and $2q$ be the relative momentum
between $c$ and $\bar c$ quarks, then $v^2 \sim 4q^2/p^2 \sim
0.25$ can be treated as a small expansion parameter \cite{BBL1}.
For P-wave charmonium $\chi_{c1}$, because the wave function at
the origin $\mathcal{R}_P(0)\!\!=\!\!0$, which corresponds to the
zeroth order in $q$, we must expand the amplitude to first order
in $q$. Thus we have
\begin{eqnarray}
 \label{amp}
\mathcal{M}(B\to\!\! \chi_{c1}K)\!=\!\!\!\!\sum_{L_z,S_z}\!\langle
1L_z;1S_z|1J_z\rangle
 \!\int\!\!\frac{\mathrm{{d}}^4 q}{(2 \pi)^3}q_\alpha\delta\!(q^0) \nonumber\\
 \times \psi_{1M}^\ast\!(q)
 \mathrm{Tr}[\mathcal{O}^\alpha\!(0)P_{1S_z}\!(p,\!0)
\!+\!\mathcal{O}\!(0)P^\alpha_{1S_z}\!(p,\!0)],
 \end{eqnarray}
where $\mathcal{O}(q)$ represent the rest of the decay amplitudes
and $P_{1S_z}(p,q)$ is the spin-triplet projection operator, and
$\mathcal{O}^\alpha,~P^\alpha$ stand for the derivatives of
$\mathcal{O},~P$ with respect to the relative momentum
$q_{\alpha}$~\cite{Chao04}.   The amplitudes $\mathcal{O}(q)$ can
be further factorized as product of $B \to K$ form factors and
hard kernel or as the convolution of a hard kernel with light-cone
wave functions of B meson and K meson, within QCD factorization
approach.

After $q^0$ is integrated out, the integral in Eq.~(\ref{amp}) is
proportional to the derivative of the P-wave wave function at the
origin by
 \bqa
\int\!\frac{\mathrm{{d}}^3 q}{(2 \pi)^3}q^\alpha \psi_{1M}^\ast(q)
=-i\varepsilon^{\ast\alpha}(L_z)\sqrt{\frac{3}{4\pi}}
\mathcal{R}^{'}_P(0), \eqa
where $\varepsilon^\alpha(L_z)$ is the polarization vector of an
angular momentum-1 system and the value of $\mathcal{R}^{'}_P(0)$
for charmonia can be found in, e.g., Ref.~\cite{Quig}.


In contrast to the NR description of $\chi_{c1}$, the  K-meson is
described by LCDAs \cite{BBNS}:
\bqa
   \langle K(p\,')|\bar s_\beta(z_2)\,d_\alpha(z_1)|0\rangle&=&\nonumber\\
   &&\hspace*{-3.8cm} \frac{i f_K}{4} \int_0^1dxe^{i(y\,p\,'\cdot z_2+\bar y \,p\,'\cdot z_1)}
\Bigl\{ \pslash{p\,'}\,\gamma_5\,\phi_K(y)\Bigr\}_{\alpha\beta},
\label{kaon}
  \eqa
where $y$ and $\bar{y}=1-y$ are the momentum fractions of the $s$
and $\bar{d}$ quarks inside the $K$ meson respectively, and
$\phi_K(x)$ is the leading twist LCDA of K-meson.  The masses of
light quarks and $K$ meson are neglected in heavy quark limit.

The effective Hamiltonian for $B \to\! \chi_{c1}K$ reads
\cite{BBL}
 \bqa
\mathcal{H}_{\mathrm{eff}}\!\!=\!\!\frac{G_F}{\sqrt{2}} \Bigl(\!
V_{cb} V_{cs}^*\!(C_1 {\cal O}_1\!+C_2 {\cal O}_2 )\!-V_{tb}
V_{ts}^* \sum_{i=3}^{6} C_i {\cal O}_i \!\Bigr),
 \eqa
where $G_F$ is the Fermi constant, $C_i$ are the Wilson
coefficients and $V_{q_1q_2}$ are the CKM matrix elements.
The relevant 4-fermion operators ${\cal O}_i$ can be found in
\cite{Chao04}.

\begin{figure}[t]
\vspace{-3.0cm}
 \hspace*{-2.2cm}
\includegraphics[width=13cm,height=16cm]{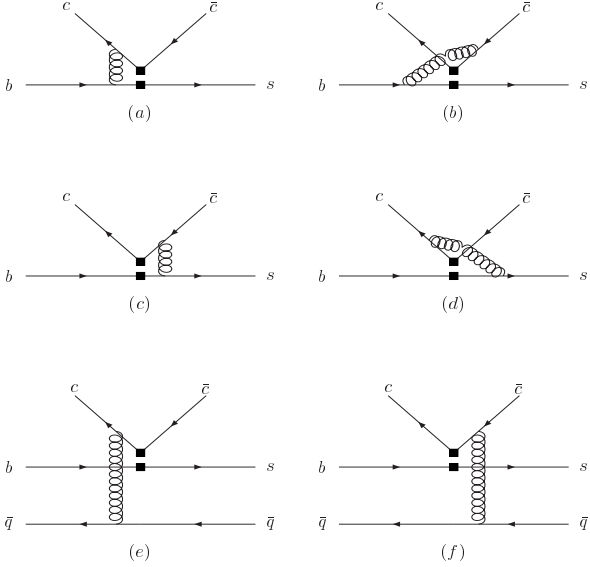}
\vspace{-5.0cm}
\caption{ Feynman diagrams for vertex and
spectator corrections to $B \to\! \chi_{c0} K$.} \label{fvs}
\end{figure}

According to \cite{BBNS} all nonfactorizable corrections are due
to Fig.\ref{fvs}.  These corrections, with  operators ${\cal O}_i$
inserted, contribute to the amplitude $\mathcal{O}(q)$ in Eq.
(\ref{amp}), where the external lines of charm and anti-charm
quarks have been truncated.  Taking nonfactorizable corrections in
Fig.\ref{fvs} into account, the decay amplitude for $B\to\!
\chi_{c1}K$ in QCD factorization is written as
\bqa\label{amp2}
  i\mathcal{M} &=&  \frac{G_F}{\sqrt{2}} \Bigl[V_{cb}
V_{cs}^* a_2 -V_{tb} V_{ts}^* (a_3 - a_5) \Bigr]\nonumber\\
   && \times 12i \sqrt{\frac{ 2  }{\pi
M}}\mathcal{R}^{'}_P(0)  \epsilon^* \cdot p_B F_1 (M^2),
 \eqa
where $\epsilon$ is polarization vector of
$\chi_{c1}$~\cite{Polar}.
 Here $F_1$ is the $B \to K$ form factor and we have used the
relation ${F_0(M^2)}/{F_1 (M^2)}=\! 1-z$ \cite{Cheng}, with
$z\!=\!M^2/m_B^2\approx 4m_c^2/m_b^2$ and $M$ is the mass of
$\chi_{c1}$, to simplify the structure of (\ref{amp2}).

The coefficients $a_i$ ($i=2,3,5$) in the naive dimension
regularization(NDR) scheme are given by
  \bqa\label{ai}
a_2\!\!\!\!&=&\!\!\!\!
C_2\!+\!\frac{C_1}{N_c}\!+\!\frac{\alpha_s}{4\pi}\frac{C_F}{N_c}
C_1 \Bigl(\!-18\! +\!12\ln \frac{m_b}{\mu} \!+\! f_I \!+ \!f_{II}
\Bigr), \nonumber \\
a_3\!\!\!\!&=&\!\!\!\!
C_3\!+\!\frac{C_4}{N_c}\!+\!\frac{\alpha_s}{4\pi}\frac{C_F}{N_c}
C_4 \Bigl(\!-18\! +\!12\ln \frac{m_b}{\mu}\!+\!f_I \!+ \!f_{II}
\Bigr),
 \nonumber\\
a_5\!\!\!\!&=&\!\!\!\!
C_5\!+\!\frac{C_6}{N_c}\!-\!\frac{\alpha_s}{4\pi}\frac{C_F}{N_c}
C_6 \Bigl(\!-6 \!+\!12\ln \frac{m_b}{\mu}\! +\!f_I\!+\!f_{II}\Bigr),
 \eqa
 where $C_F=(N_c^2-1)/(2 N_c)$ and $\mu$ is the  QCD renormalization scale.

 The function $f_I$ is calculated from the four vertex correction
diagrams (a, b, c, d) in Fig.\ref{fvs} and reads
  \bqa
  f_I\!&=&\!\frac{2\,z}{2 - z} - \frac{4\,z\,\log (4)}{2 - z} -
  \frac{4\,z^2\,\log (z)}{\left( 1 - z \right) \,\left( 2 - z \right) }\nonumber\\
  && + \frac{4\,\left( 3 - 2\,z \right) \,\left( 1 - z \right) \,
     \left( \log (1 - z)-i\,\pi \right) }{{\left( 2 - z \right) }^2},
 \eqa
We find that the infrared divergences are canceled between
diagrams (a) and (b), (c) and (d) respectively in Fig.\ref{fvs}.
On the other hand, this function is different from that in
Eq.~(11) of Ref.~\cite{Chao03} even when a nonrelativistic limit
wave function $\phi_{\chi_{c1}}^{NR} (u)\! =\! \delta (u-1/2) $ is
adopted, as we have mentioned.

For the two spectator correction diagrams (e,\! f) in
Fig.\ref{fvs}, the off-shellness of the gluon is natural to be
associated with a scale $\mu_h \sim \sqrt{m_b
\Lambda_{\mathrm{QCD}}}$, rather than $\mu_h \sim m_b $. Following
Ref.~\cite{BBNS}, we choose $\mu=\sqrt{m_b \Lambda_h}\approx 1.4$
Gev with $\Lambda_h=0.5$ Gev in calculating the hard spectator
function $f_{II}$ and then, in the leading twist approximation, we
get
 \bqa\label{fII}
f_{II}&=&\frac{\as(\mu_h ) C_i(\mu_h)}{\as(\mu )
C_i(\mu)}\frac{8\pi^2}{N_c} \frac{f_K f_B}{F_1 (M^2)
m_B^2} \frac{1}{1-z}\nonumber\\
&&\hspace*{-1.0cm} \times\int_0^1 d\xi \frac{\phi_B (\xi)}{\xi}
\int_0^1 dy \frac{\phi_K (y)}{y}[1+\frac{z}{y(1-z)}],
 \eqa
  where $\xi$ is the momentum fraction of the
spectator quark in the $B$ meson and $C_i(\mu_h)$ ($i=1,4,6$) are
the NLO Wilson coefficients which can be evaluated by the
renormalization group approach \cite{BBL}.

The spectator contribution depends on the wave function $\phi_B$
through the integral
 \bqa
\int_0^1 d\xi \frac{\phi_B(\xi)}{\xi} \equiv
\frac{m_B}{\lambda_B}.
 \eqa
 Since $\phi_B (\xi)$ is appreciable only for $\xi$ of order
$\Lambda_{\mathrm{QCD}}/m_B$, $\lambda_B$ is of order
$\Lambda_{\mathrm{QCD}}$. We will choose $\lambda_B\approx 300$
MeV in the numerical calculations \cite {BBNS}.

 If we choose the asymptotic form of the K meson twist-2 LCDA ,
$ \phi_K (y)=6y(1-y)$, we can find logarithmic end-point
singularities in Eq.~(\ref{fII}) just like that in
Ref.~\cite{Chao03}, and we parameterize it in a simple way,
\begin{equation}\label{endpoint}
\int \frac{dy}{y}=\ln \frac{m_B}{\Lambda_h}\approx 2.4.
\end{equation}


The mass of $\chi_{c1}(1P)$ $M_{\chi_{c1}}\!\!\!=\!\!3.511$ Gev is
known, but the mass of the missing charmonium $\chi_{c1}(2P)$ has
to be estimated by, say, potential models.  We choose
$M_{\chi_{c1}'}\!\!\!=\!\!3.953$ Gev following
Ref.~\cite{Godfrey}. Then the form factor $F_1(M^2)$ can be
determined by light-cone sum rules~\cite{Ball},
 \bqa \label{F1}
 F_1(M_{\chi_{c1}}^2)=0.80,~~F_1(M_{\chi_{c1}'}^2)=1.14.
 \eqa
We also choose $M_{\chi_{c1}'}=3.872$ Gev and
$F_1(M_{\chi_{c1}'}^2)=1.06$ to study  if  the $X(3872)$ behaves
like a $\chi_{c1}(2P)$ in their b-production processes.

For numerical analysis, we use the following input parameters :
 \bqa \label{parameter}
&&\!\!  m_b
=4.8 ~\mbox{GeV}, ~ \!\! m_B=5.28~\mbox{GeV},~\!\! \!f_K\!=\!160~\mbox{MeV}, \nonumber \\
&& \!\!f_B\!=\!216 ~\mbox{MeV\cite{Gray}},
~\!\!\mathcal{R}^{'}_{1P}(0)\!=\!\mathcal{R}^{'}_{2P}(0)\!=\!\sqrt{0.1}~\!\mbox{GeV}^{5/2},\nonumber \\
&&\!\!C_1(\mu)=1.21(1.082),~
C_2(\mu)=-0.40(-0.185),\nonumber\\
&&\!\!C_3(\mu)=0.03(0.014),~
C_4(\mu)=-0.05(-0.035),\nonumber\\
&&\!\!C_5(\mu)=0.01(0.009),~
C_6(\mu)=-0.07(-0.041),\nonumber\\
&& \!\!\as(\mu)=0.35(0.22).
 \eqa
In (\ref{parameter}) the $\mu$-dependent quantities at
$\mu_h\!=\!1.4$ Gev ($\mu\!=\!4.4$ Gev) are shown without (with)
parentheses.

\begin{table}[tb]
\begin {center}
\begin{tabular}{ c|ccc}
 \hline
 $$               &$a_2$       &$a_3$        &$a_5$ \\   \hline
$\chi_{c1}(3511) $&0.199-0.051i& 0.000+0.002i&0.004-0.002i \\
$\chi_{c1}'(3953)$&0.247-0.042i&-0.002+0.001i&0.007-0.002i \\
$\chi_{c1}'(3872)$&0.236-0.044i&-0.002+0.001i&0.006-0.002i \\
\hline
\end{tabular}
\caption{The coefficients $a_i$ of $B\to\chi_{c1}(1P,2P)K$ with
different choices of $M_{\chi_{c1}'}$.}
 \label{table1}
\end {center}
\vspace{-0.5cm}
\end{table}

Using the above inputs, we get the results of coefficients $a_i$
which are listed in Table.~\ref {table1}.  With the help of these
coefficients $a_i$, we calculate the decay branching ratios of
decays $B\to\chi_{c1}(1P,2P)K$ with two different choices of
$M_{\chi_{c1}'}$ and get
 \bqa\label{branching}
{\mathrm{Br}} (B^0 \rightarrow \chi_{c1}(3511) K^0) &=&
1.79 \times 10^{-4},\nonumber\\
{\mathrm{Br}} (B^0 \rightarrow \chi_{c1}'(3953) K^0) &=&
1.81 \times 10^{-4}, \nonumber\\
{\mathrm{Br}} (B^0 \rightarrow \chi_{c1}'(3872) K^0) &=&
 1.78 \times 10^{-4}.
 \eqa

 Our prediction of ${\mathrm{Br}}
(B^0 \rightarrow \chi_{c1}(3511) K^0)$ is about 2 times larger
than that in \cite{Chao03}  , although it is still about two times
smaller than the recent data \cite{BaBar05}.
The difference between the theoretical predictions and
experimental data may not be as serious as it looks like if we
take into account the following uncertainties: (i) We have used a
moderate value of $\mathcal{R}^{'}_{1P}(0)$ predicted by different
potential models \cite{Quig} in our calculation, and a larger
value of $\mathcal{R}^{'}_{1P}(0)$ may enhance our prediction in
Eq.~(\ref{branching}) significantly.  (ii) In evaluation of
$f_{II}$, we only use the leading twist LCDAs of K-meson, and
large uncertainties will arise from the chirally enhanced higher
twist effects \cite{Cheng}.  (iii) Since the squared velocity
$v^2$ of the charm quark in charmonium is about 0.25-0.30, the
relativistic corrections may be important for these decays.

Note that although the form factor in (\ref{F1}) and the
coefficient $a_2$ in Table.~\ref {table1} increase evidently as
the charmonium mass increases, the decreased phase space and
kinematic factors in (\ref{amp2}) will make a balance, and result
in similar decay branching ratios in the charmonium mass region
3.51-3.95~GeV, as shown in (\ref{branching}). If we neglect the
order $\as$ corrections (i.e., in the naive
factorization~\cite{BSW}), the ratios among these three branching
fractions in (\ref{branching}) would become 1~:~0.74~:~0.69. As
estimated in (\ref{branching}), the branching ratios for
$\chi_{c1}(2P)$ are
 \bqa\label{chic2p}
{\mathrm{Br}} (B^0 \!\rightarrow \!\chi_{c1}' K^0)\! &\approx &\!
2\times 10^{-4},\nonumber\\
{\mathrm{Br}} (B^+ \!\rightarrow \!\chi_{c1}' K^+)\!
&=&\!{\mathrm{Br}} (B^0 \!\rightarrow \!\chi_{c1}' K^0).
 \eqa

Comparing Eq.~(\ref{chic2p}) with the measured channel of the
$X(3872)$ \cite{Belle03}:
 \bqa\label{psipipi}
{\mathrm{Br}} (B^+ &\!\rightarrow& \!X K^+)\times\mathcal{B}_X
 = (1.3\pm 0.3)\times 10^{-5},\\
 \mathcal{B}_X &\equiv& {\mathrm{Br}} (X\!\rightarrow\! J/\psi \pi^+\pi^-),\nonumber
 \eqa
we see that the produced X(3872) looks like the $\chi_{c1}(2P)$ if
$\mathcal{B}_X$ is sufficient small, say, $3\sim7\%$. A similar
conclusion has recently been obtained in a comprehensive analysis
of X(3872) production at the Tevatron and B-factories
\cite{Bauer}. On the other hand, if X(3872) is a loosely bound
S-wave molecule of $D^0\bar{D}^{*0}/D^{*0}\bar{D}^0$
\cite{Tornqvist,Swanson}, a model calculation gives a smaller rate
\cite{Braaten} compared with Eq.~(\ref{chic2p}):
 \bqa\label{braaten1}
{\mathrm{Br}} (B^+ &\!\rightarrow& \!X K^+)
 = (0.07\sim 1)\times 10^{-4},
 \eqa
which requires a larger $\mathcal{B}_X > 10\%$ to be consistent
with the experimental data (\ref{psipipi}). They also predict:
\bqa\label{braaten2}
 Br (B^0\!\!  \rightarrow \!   X(3872) K^0) <
0.1Br (B^+ \!\! \rightarrow
 \! X(3872) K^+).
 \eqa
So the measurement of $\mathcal{B}_X$ and $Br (B^0\!  \rightarrow
\!   X(3872) K^0)$ is very helpful to identify the nature of
X(3872).

Recently, a preliminary result for a new decay mode $X\rightarrow
D^0\bar D^0\pi^0$ was found by Belle\cite{Olsen}:
 \bqa\label{ddpi}
{\mathrm{Br}} (B\!&\rightarrow&\!XK)\times{\mathrm{Br}}
(X\rightarrow D^0\bar D^0\pi^0)\nonumber\\
 &=& (2.2\pm 0.7\pm 0.4)\times 10^{-4}.
 \eqa
Eq.~(\ref{ddpi}) implies that $\mathcal{B}_X < 10\%$, if it can be
confirmed by further measurements.  This would disfavor the
suggestion that the X(3872) is a loosely bound S-wave molecule of
$D^0\bar{D}^{*0}/D^{*0}\bar{D}^0$ with predictions of both
decay\cite{Swanson} and production\cite{Braaten}.

The above discussions about the X(3872) is based on the assumption
that the X(3872) is a pure charmonium $\chi_{c1}(2P)$ state. But
this cannot be the case due to the coupled channel effects and
X(3872) being in extremely close proximity to the
$D^0\bar{D}^{*0}/D^{*0}\bar{D}^0$ threshold. Perhaps a more
realistic model for the X(3872) (for further discussions
see~\cite{chao}) is that the X(3872) has a dominant
 $J^{PC} = 1^{++}(2P)$ $c\bar c$ component which is mixed with a
substantial real $D^0\bar{D}^{*0}/D^{*0}\bar{D}^0$ continuum
component (the $D^+\bar{D}^{*-}/D^{*-}\bar{D}^+$ continuum
component is kinematically forbidden to be mixed in X(3872) and it
is the $u-d$ quark mass difference that causes this isospin
violation). Thus X(3872) will have the following features. (1)~The
production of X(3872) in B meson decays is mainly due to the
$J^{PC} = 1^{++}(2P)$ $c\bar c$ component, as discussed above. The
production of X(3872) at the Tevatron is also due to this $c\bar
c$ component and associated higher Fock states containing the
color-octet $c\bar c$ pair and soft gluons. As was
argued~\cite{chao1} for the prompt charmonium production that
cross sections of D-wave charmonia (which were suggested as a
tentative candidates for X(3872) in~\cite{chao1}) could be as
large as $J/\psi$ or $\psi(2S)$ due to the color-octet mechanism,
the P-wave $(2P)$ charmonium could also have the comparable
production rate to $J/\psi$ or $\psi(2S)$. But this does not seem
to be obvious for a loosely bound S-wave molecule of
$D^0\bar{D}^{*0}/D^{*0}\bar{D}^0$. (2)~On the other hand, the
$D^0\bar{D}^{*0}/D^{*0}\bar{D}^0$ continuum component in X(3872)
will be mainly in charge of the hadronic decays of X(3872) into
$D^0\bar{D}^{*0}/D^{*0}\bar{D}^0$ or $D^0\bar D^0\pi^0$ as well as
$J/\psi\rho^0$ and $J/\psi\omega$. The latter two decay modes
($J/\psi\rho^0$ and $J/\psi\omega$) may come from the first decay
mode $D^0\bar{D}^{*0}/D^{*0}\bar{D}^0$ and a subsequent
rescattering final state interaction and therefore have the same
decay amplitudes [A($J/\psi\rho^0$)=A($J/\psi\omega$)] that are
smaller than the first decay mode amplitude. (3)~A substantial
$D^0\bar{D}^{*0}/D^{*0}\bar{D}^0$ continuum component in X(3872)
may reduce the production rates in Eq.~(\ref{chic2p}), and will
also reduce the $X(3872)\rightarrow J/\psi\gamma$ decay width,
which can be as small as $11$~KeV~\cite{BG} (note that this 2P-1S
E1 transition is sensitive to the model details, see,
e.g.~\cite{Swanson}). This is much smaller then the hadronic decay
widths. But a large rate for
$\chi_{c1}(2P)\rightarrow\gamma\psi(2S)$=60-100 KeV will be
expected. These qualitative features are useful in understanding
the nature of X(3872) and should be further tested and studied
experimentally and theoretically.


In summary, we study the decays $B\to\chi_{c1}(1P,2P)K$ in QCD
factorization by treating charmonia as nonrelativistic bound
states.  We find that there are no infrared divergences in the
vertex corrections, and the logarithmic end-point singularities
from  hard spectator interactions can be regularized by a momentum
cutoff. Within certain uncertainties we find the
$B\to\chi_{c1}(2P)K$ decay rate can be comparable to
$B\to\chi_{c1}(1P)K$, and get $Br(B^0 \!\rightarrow \!\chi_{c1}'
K^0)\! =\! Br(B^+ \!\rightarrow \!\chi_{c1}' K^+)\! \approx \!
2\times 10^{-4}$. This might imply that the X(3872) has a dominant
$J^{PC} = 1^{++}(2P)$ $c\bar c$ component but mixed with some
$D^0\bar{D}^{*0}+D^{*0}\bar{D}^0$ continuum component. The
qualitative features of X(3872) are discussed and should be
further tested and studied.

$Note$. After this work appeared in hep-ph/0506222 we learned some
new results from BaBar~\cite{BaBar05b}:
${\mathrm{Br}} (B^+ \!\rightarrow \!X(3872) K^+)
 < 3.2~\times 10^{-4},
R=\frac{Br (B^0  \rightarrow   X(3872) K^0)}{Br (B^+  \rightarrow
 X(3872) K^+)} = 0.50\pm 0.30\pm 0.05.$
 We also note that a recent paper\cite{suzuki} (hep-ph/0508258) obtained similar
 conclusions to ours for the X(3872).

\begin{acknowledgments}
We thank G. Bauer, D. Bernard, S. Olsen, and V. Paradimitriou for
stimulating discussions on the experimental status of the X(3872),
and E. Swanson for helpful comments. This work was supported in
part by the National Natural Science Foundation of China (No
10421003), and the Key Grant Project of Chinese Ministry of
Education (No 305001).

\end{acknowledgments}



\begin{thebibliography}{}
\bibitem{BSW}
M.~Bauer, B.~Stech and M. Wirbel, Z. Phy. C {\bf 34}, 103 (1987).


\bibitem{Chao03}
 Z.Z. Song and K.T. Chao, Phys. Lett.  B
{\bf 568 }, 127 (2003).

\bibitem{BBNS}
M. Beneke, G. Buchalla, M. Neubert and C.~T. Sachrajda, Phys.\
Rev.\ Lett. {\bf 83 }, 1914 (1999); Nucl. Phys. B {\bf 591 }, 313
(2000); Nucl. Phys. B  {\bf 606 }, 245 (2001).

\bibitem{Li}
 C. H. Chen  and H. N. Li,
hep-ph/0504020.

\bibitem{Chao04}
Z.~Z. Song, C.~Meng, Y.~J. Gao and K.~T. Chao,  Phys.\ Rev.\ D
{\bf 69}, 054009 (2004).

\bibitem{Chao02}
Z.~Z. Song, C.~Meng and K.~T. Chao, Eur.\ Phys.\ J.\ C {\bf 36},
365 (2004)


\bibitem{BBL1}
 G.~T. Bodwin, E. Braaten and G.~P. Lepage, Phys.\ Rev.\ D {\bf 51
 }, 1125 (1995);  {\bf 55}, 5853(E) (1997).




\bibitem{Belle03}
S. K. Choi {\it et al.} (Belle Collaboration), Phys.\ Rev.\ Lett.\
{\bf 91}, 262001 (2003); D. Acosta {\it et al.} (CDF II
Collaboration), Phys.\ Rev.\ Lett.\ {\bf 93}, 072001 (2004); V. M.
Abazov {\it et al.} (D0 Collaboration), Phys.\ Rev.\ Lett.\ {\bf
93}, 162002 (2004) ; B.~Aubert {\it et al.} (BaBar Collaboration),
hep-ex/0406022.

\bibitem{Belle05}
K. Abe {\it et al.} (Belle Collaboration), hep-ex/0505038.

\bibitem{BG}
T. Barnes and S. Godfrey, Phys. Rev. D 69, 054008 (2004); E.J.
Eichten, K. Lane and C. Quigg, Phys. Rev. D69, 094019 (2004).

\bibitem{chao1} K.T. Chao, Prompt D-wave charmonium production at the Tevatron and the X(3870),
talk given at the Second Workshop of the Quarkonium Working Group,
Fermilab, Sep.20-22, 2003, [{\rm
http://www.qwg.to.infn.it/WS-sep03/WS2talks/prod/chao.ppt}].

\bibitem{Tornqvist}
 N.A. Tornqvist, Phys. Lett. B
{\bf 590}, 209(2004); F. Close and P. Page, Phys. Lett. B578,
119(2004); C.Y. Wong, Phys.Rev. C69, 055202(2004); M.B. Voloshin,
Phys.Lett. B604, 69(2004); M.T. AlFiky et al., hep-ph/0506141.

\bibitem{Swanson}
E. Swanson, Phys.Lett. B{\bf 588},189(2004); {\bf 598},197(2004).

\bibitem{Braaten}
E. Braaten, M. Kusunoki and S. Nussinov , Phys.\ Rev.\ Lett.\ {\bf
93}, 162001 (2004);E. Braaten and M. Kusunoki Phys.\ Rev.\ D {\bf
71}, 074005 (2005).

\bibitem{Quig}
 E.J. Eichten and C. Quigg, Phys.\ Rev.\ D {\bf 52}, 1726 (1995).



\bibitem{BBL}
 G. Buchalla, A.~J. Buras and M.~E. Lautenbacher,
Rev. Mod. Phys. {\bf 68}, 1125 (1996).
\bibitem{Polar}
 The polarization vector $\epsilon$ comes from the tensor-sum of orbit polarization and
 total spin. For details, see \cite{Chao04}.
\bibitem{Cheng}
H.~Y. Cheng and K.~C. Yang, Phys.\ Rev.\ D {\bf 63}, 074011(2001);
J. Chay and C. Kim, hep-ph/0009244.

\bibitem{Godfrey}
 S. Godfrey and N. Isgur, Phys.\ Rev.\ D {\bf 32}, 189 (1985).

\bibitem{Ball}
 P. Ball and R. Zwichy,
hep-ph/0406261.

\bibitem{Gray}
 A. Gray {\it et al.},
hep-lat/0507015.


\bibitem{BaBar05}
B.~Aubert {\it et al.} (BaBar Collaboration), Phys.\ Rev.\ Lett.\
{\bf 94}, 141801 (2005); Phys.\ Rev.\ D {\bf 65}, 032001 (2002).


\bibitem{Bauer}
G. Bauer, hep-ex/0505083.



\bibitem{Olsen}
S. L. Olsen, talk given at the APS/DPF Meeting 2005, [{\rm
http://belle.kek.jp/belle/talks/aps05/olsen.pdf}].

\bibitem{chao} K.T. Chao, talk given at the Workshop on New Hadron States, Beijing, October 2005.





\bibitem{BaBar05b}
 B.~Aubert {\it et al.} (BaBar Collaboration), hep-ex/0507090, hep-ex/0510070.

\bibitem{suzuki}
M. Suzuki, hep-ph/0508258.

\end{thebibliography}
\end{document}